\documentclass[twocolumn,showpacs,preprintnumbers,amsmath,amssymb]{revtex4}
\usepackage{graphicx}% Include figure files
\usepackage{dcolumn}% Align table columns on decimal point
\usepackage{bm}% bold math

\newcommand\T{\rule{0pt}{2.6ex}}
\newcommand\B{\rule[-1.2ex]{0pt}{0pt}}

\begin{document}

%opening
\title{Computer simulation of fatigue under diametrical compression}
\author{H.\ A.\ Carmona$^{1,2}$, F.\ Kun$^3$, J.\ S.\ Andrade Jr.$^4$, and
H.\ J.\ Herrmann$^2$}
\affiliation{$^1$Centro de Ci\^encias e Tecnologia, Universidade Estadual do Cear\'a, 
60740-903 Fortaleza, Cear\'a, Brazil}
\affiliation{$^2$IfB, HIF, E18, ETH, H\"onggerberg, 8093 Z\"urich, Switzerland}
\affiliation{$^3$Department of Theoretical Physics, University of
Debrecen, P. O. Box:5, H-4010 Debrecen, Hungary} 
\affiliation{$^4$Departamento de F\'isica, Universidade Federal do
Cear\'a, 60451-970 Fortaleza, Cear\'a, Brazil} 

\begin{abstract}
We study the fatigue fracture of disordered materials by means of
computer simulations of a discrete element model. We extend a
two-dimensional fracture model to capture the microscopic mechanisms
relevant for fatigue, and we simulate the
diametric compression of a disc shape specimen under a constant
external force.
The model allows to follow the development of the fracture
process on the macro- and micro-level varying the relative influence
of the mechanisms of damage accumulation over the load history and
healing of microcracks. As a specific example we consider recent
experimental results on the fatigue fracture of asphalt. Our numerical
simulations show that for intermediate applied loads the lifetime of
the specimen presents a power law behavior. Under the effect of
healing, more prominent for small loads compared to the tensile
strength of the material, the lifetime of the sample increases and a
fatigue limit  emerges below which no macroscopic failure occurs. 
The numerical results are in a good qualitative agreement with the
experimental findings.
\end{abstract}

\maketitle

\section{Introduction}
Understanding the fatigue damage process of disordered materials is
obviously important in many areas. The time scales related to the failure process, 
the relevant damage mechanisms, and scaling laws associated
with this phenomenon have been widely studied both experimentally
and theoretically \cite{Zapperi:2006,hans:book,Shcherbakov:2003,Turcotte:1992}. 
Universal laws have been found in the fracture process related to both
system sizes and geometry \cite{hans:77,Carpinteri:2005},  
and in particular damage accumulation has been shown to play an
important role in the formation and development of cracks with fractal
structure   
\cite{hans:086}.

This phenomenon is also important in practical applications, in
particular when dealing with asphalt mixtures subject to traffic
loading. Fatigue 
cracking is one of the main causes for asphalt layer failure in
pavement structures, among moisture damage and thermal cracking. Most
of the knowledge about fatigue failure of asphalt mixtures relies,
however, on experimental observations \cite{matthews:1993,
daniel:2001, si:2002, kim:2003, lundstrom:2004, castro:2006}. Material
modeling to improve structural design is becoming an important tool to
overcome this mode of material distress. 

Recently, fatigue life tests have been carried out to study the performance of
asphalt mixtures \cite{hans:424}, by measuring the accumulation of
deformation with time for cylindrical discs under diametrical
compression applied periodically with a constant
amplitude. Experiments revealed that the fatigue process has three
different regimes depending on the external load amplitude $\sigma$:
when $\sigma$ falls close to the tensile strength of the material a
rapid failure occurs, while for low load values a so-called fatigue
limit emerges below which the material does not suffer macroscopic
breaking. In the intermediate load regime the lifetime of the specimen
$t_f$ has a power law dependence on the load, also called the Basquin
law of fatigue \cite{basquin:1910}.
Different models have been developed to obtain a theoretical
understanding of the fatigue performance of asphalt specimens \cite{kunPRL:r3,
kunPRL:r4,kim:2004}. 
Besides damage accumulation, experimental research has shown the
importance of the healing process to the lifetime of asphalt mixtures
\cite{kim:1991, daniel:2001, si:2002, kim:2003, castro:2006}. The
healing mechanism is related to the recovery of microcracks due to the
viscoelastic nature of the binder material in polymeric mixtures,
resulting in an extended lifetime. Theoretical approaches turned out
to have difficulties in capturing all the mechanisms involved in the
fatigue process, such as damage growth, relaxation due to
viscoelasticity and healing \cite{si:2002, kunPRL:r3, kunPRL:r4,
kunPRL:r5, kunPRL:r1}. 
Although, these models agree well with the results                    
of the experiments, taking into account the stochastic nature of the
fracture process and the cumulative effect of the load history, they
fail to describe explicitly the processes of damage, fracture and
failure of the solid. A more complete understanding of the failure
process due to damage accumulation in the Brazilian test configuration
may benefit from detailed computer simulations  confronted with the
experiments, as already was successfully performed in many fracture
processes \cite{hans:188, hans:198, hans:237, hans:263, vanMier:1999,
tang:2006, mariotti:2001, thornton:2004, vandeSteen:2005}. 

In this paper we enhance a 2D discrete element model of the fracture
of disordered materials in order to capture the relevant microscopic
mechanisms of fatigue. 
In the model convex polygons symbolizing grains
of the material are coupled by breakable beams, which can suffer
stretching and bending deformation
\cite{hans:68, hans:book, hans:188,hans:237,hans:263, Hansen:2001,Timonen:2004,Timonen:2005,Daddetta:2006,Zapperi:2006, astrom:2006}. 
The beams of the model can
represent the polymeric binder of asphalt between aggregates.
Similarly to former studies, breaking of a beam can be caused by stretching and bending
captured by the von Mises-type breaking criterion
\cite{hans:188,hans:237,hans:263, Daddetta:2006}. Additionally, we 
introduce an ageing mechanism, i.e.\ intact beams undergo a damage
accumulation process which can again give rise to breaking. The
accumulation process introduces memory over the loading history of the
system. Damage recovery leading to healing is implemented in the
model by limiting the range of memory which contributes to the amount
of damage of fibers. 
We study in details the time evolution of the
system and demonstrate that the model provides a good quantitative
description of the experimental findings, where damage
accumulation and healing proved to be essential.

\section{Model}
Our model is an extension of a realistic discrete element model of
disordered materials which has been successfully applied to study
various aspects of fracture and fragmentation phenomena
\cite{hans:188,hans:237,hans:263,Daddetta:2006}. 
The material is assumed to be composed of a large number of mesoscopic
elements that interact  elastically with each other. In our model the mesoscopic
elements are convex polygons randomly generated using a Voronoi
construction \cite{hans:267}. 

The mesoscopic elements are elastic and unbreakable bodies, all with
the same physical properties. Deformations are represented in the
model by the possible overlap of the elements when pressed against
each other. Between overlapping polygons we introduce a repulsive
force which is proportional to the element's bulk Young
modulus $Y$ and the overlapping area, and has a direction perpendicular
to the contact line \cite{hans:263, hans:267}.

\begin{figure}
	\centering
	\includegraphics[width=8.2 cm,bb=0 0 555 499]{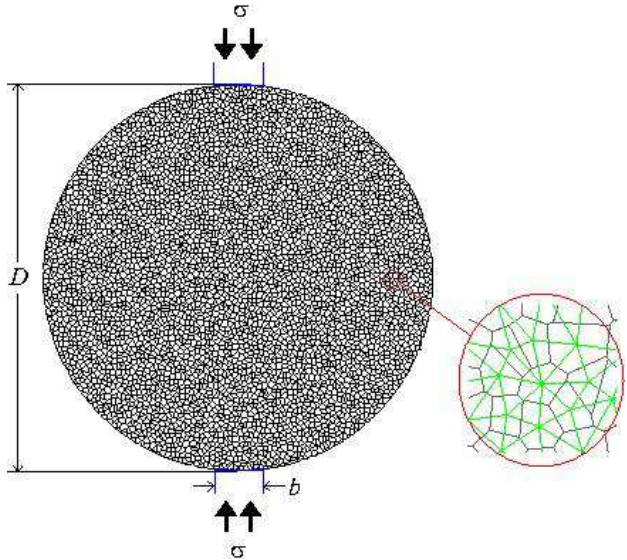}
	\caption{{\it (Color online)} Geometrical layout of the
model. A disk shaped sample is subjected to diametrical
compression. The disk is composed of convex polygons connected by
elastic beams (magnified image).}  
	\label{geometry}
\end{figure}

Cohesion between the elements is introduced in the model by the
inclusion of beams between neighboring polygons, see Fig.\
\ref{geometry}. 
The use of beams, as opposed to simple springs, takes into account tension and bending deformations, which are
especially important in the complex multi-axis stress field regime present
in the geometries used experimentally.  This model also resolves the local
torques and shear forces which naturally arise when dealing with materials with
complex micro-structures, where grains or aggregates are embedded in a matrix
material. Beams have been particularly used in the literature to prevent an unphysical failure of the system 
under shear \cite{Daddetta:2006}, which happens when using central force elements. A detailed description of 
the beam model is given elsewhere \cite{hans:68, hans:263, hans:267}. Briefly, the centers of mass of
neighboring elements are connected by beams which can suffer
stretching and bending deformation, and hence, exert an
elastic force and torque on the polygons when stretched or bent 
\cite{hans:68, hans:book, hans:188,hans:237, Hansen:2001,Timonen:2004,Timonen:2005,
Daddetta:2006,Zapperi:2006, astrom:2006}.

\begin{table}
\caption{Parameter values used in the simulations.}
\centering
\begin{tabular}{p{4.7cm}l p{2.7cm}l p{2.7cm}l}
\hline
\T Parameter  \B &  &  Value \\
\hline

\T 
Number of elements &  & $5070$ \\

Density	& $\rho$\hspace{0.5 cm} &  $5$ $g/cm^3$  \\
Bulk Young modulus & $Y$ &  $1 \times 10^{10}$ $dyn/cm^2$ \\
Beam Young modulus & $E$ &  $5 \times 10^{10}$ $dyn/cm^2$ \\

Time step & $\delta t$ &  $1 \times 10^{-6}$ $s$  \\
Diameter of the disk & $D$ &  $20$ $cm$ \\
Typical size of a single element & & $0.5$ $cm$\\
Width of the load platen & $b$ & $2.5$ $cm$ \\

Memory factor & $f_0$ &  $10-500$  \\
Range of memory& $\tau$  &   $500-\infty$ \B \\		
\hline
\end{tabular}
\label{tab:params}
\end{table}

In the model beams can break according to specified rules in order to explicitly model damage, fracture and
failure of the solid \cite{hans:263,hans:267}. 
The imposed breaking rule must take into account both the immediate
breaking of the beams by stretching and bending, as well as the
accumulation of damage and the healing mechanism. In order to achieve
these conditions for each beam, at time $t$, we evaluate the
quantities $p(t)$ and $q(t)$ defined by 
\begin{subequations}\label{pval}
\begin{align}
p(t)& = \left(\frac{\varepsilon}{\varepsilon_{th}}\right)^2 + \frac{\text{max}\left( |\theta_i|, |\theta_j| \right) }{\theta_{th}}, \label{pvala}\\
q(t)& = p(t) + f_0 \int_{0}^{t} e^{\frac{-\left( t - t'\right)}{\tau}}p(t')dt',\label{pvalb}
\end{align}
\end{subequations}
where $\varepsilon = \Delta l/l_0$ is the longitudinal deformation of the
beam, and 
$\theta_i$ and $\theta_j$ are the rotation angles at the ends of beams
between sites $i$ and $j$, respectively. Equation (\ref{pvala}) has the
form of the von Mises plasticity criterion describing the mechanical
strength of beams with respect to stretching and bending deformation. 
The first part of Eq.\ (\ref{pvala}) refers to the
breaking of the beam through stretching and the second through
bending, with $\varepsilon_{th}$ and $\theta_{th}$ being the threshold
values for elongation and bending, respectively. Equation (\ref{pvalb})
refers to the long term memory, and accounts for the accumulation of
damage and the healing mechanism. 
The rate of damage accumulation is assumed to have the form $\Delta
c(t) = f_0p(t)$, i.e.\ the damage accumulated until time $t$ is
obtained as an integral over the loading history of elements $c(t) =
f_0\int_0^tp(t')dt'$. In our model healing is captured such that
microcracks nucleated in beams can recover, a process which limits the total amount
of damage. We describe this effect by introducing a finite range of
memory over which the load experienced by beams has a contribution to
the damage. For polymeric materials the dynamics of long chain
molecules typically lead to an exponential form of the healing term. 
In Eq.\ (\ref{pvalb}) the parameter
$f_0$ is a memory factor and $\tau$ is the time range of the memory
over which 
the loading history of the specimen contributes to the accumulation of
damage \cite{hans:086}, and therefore a parameter that controls the
healing mechanism.  The breaking condition Eqs. (\ref{pvala}) and 
(\ref{pvalb}) are evaluated at each iteration time step and those beams
for which $q(t) \geq 1$ are considered to be broken, i.e., are removed from the simulation. 

For simplicity, all the beams have the same threshold values $\varepsilon_{th}$ and $\theta_{th}$, 
and disorder is introduced in the model solely through the mesh generation
\cite{hans:188,hans:198,hans:237,hans:263,hans:267,DAddetta:phd}. The global material properties can be tuned by
adjusting geometrical and microscopic physical parameters of the
model. The randomness of the 
Voronoi tessellation and the average size of the mesoscopic elements,
which define the average length and width of the beams are geometrical
parameters that, together with the Young modulus of polygons $Y$ and
that of the beams $E$ determine the macroscopic response of the model
material \cite{hans:188,hans:198,hans:237,hans:263,hans:267,hans:357}.  
The time evolution of the system is calculated by numerically solving
the equations of motion for the translation and rotation of all
elements using a sixth-order predictor-corrector algorithm. 
The breaking rule is evaluated at each iteration step.
The breaking of beams is irreversible, which means that it is
excluded from the force calculations for all following time steps. 
Table \ref{tab:params} summarizes the parameter values used in the
simulations.

\begin{figure}
\begin{center}
$\begin{array}{c@{\hspace{-0.2 cm}}c}
	\multicolumn{1}{l}{\mbox{\begin{large}(a)\end{large}}} &
 	\multicolumn{1}{l}{\mbox{\begin{large}(b)\end{large}}}  \\ [0.2cm]
	\includegraphics[width=4.4cm,bb=0 0 555 555]{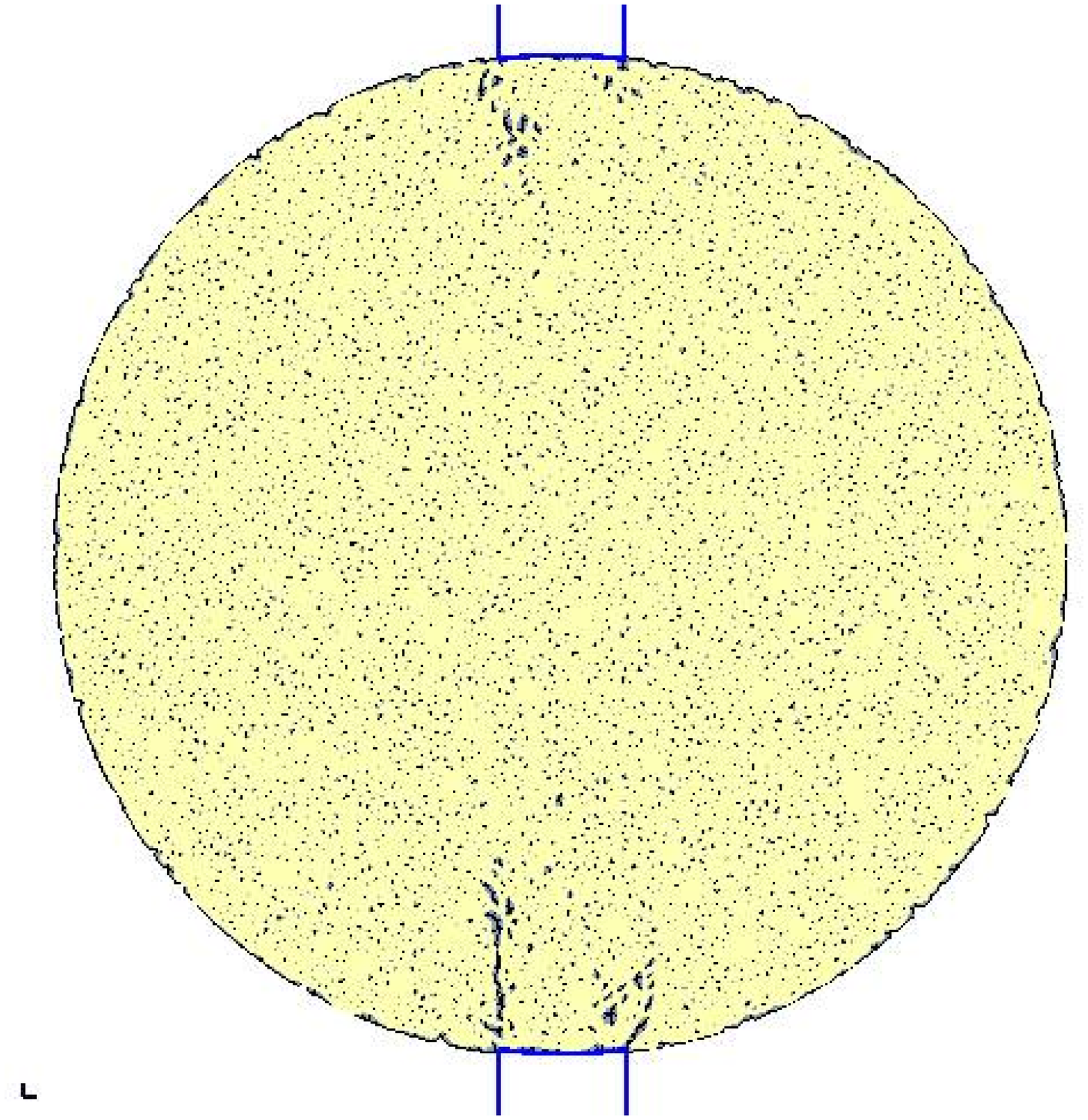}   &
	\includegraphics[width=4.4cm,bb=0 0 555 555]{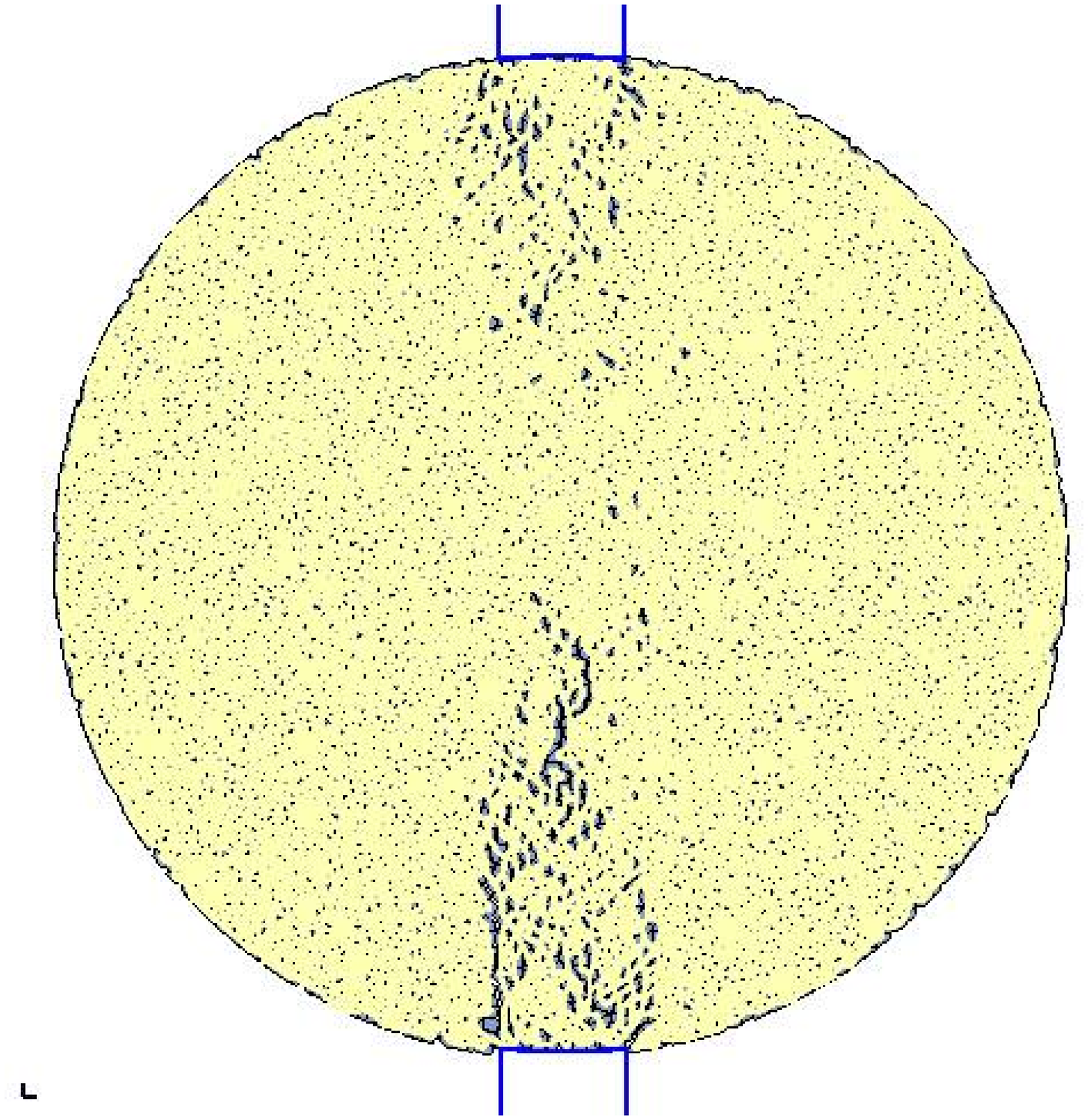} \\ [0.0cm] %t = 213

	\multicolumn{1}{l}{\mbox{\begin{large}(c)\end{large}}} &
	\multicolumn{1}{l}{\mbox{\begin{large}(d)\end{large}}} \\ [0.2cm]
	\includegraphics[width=4.4cm,bb=0 0 555 555]{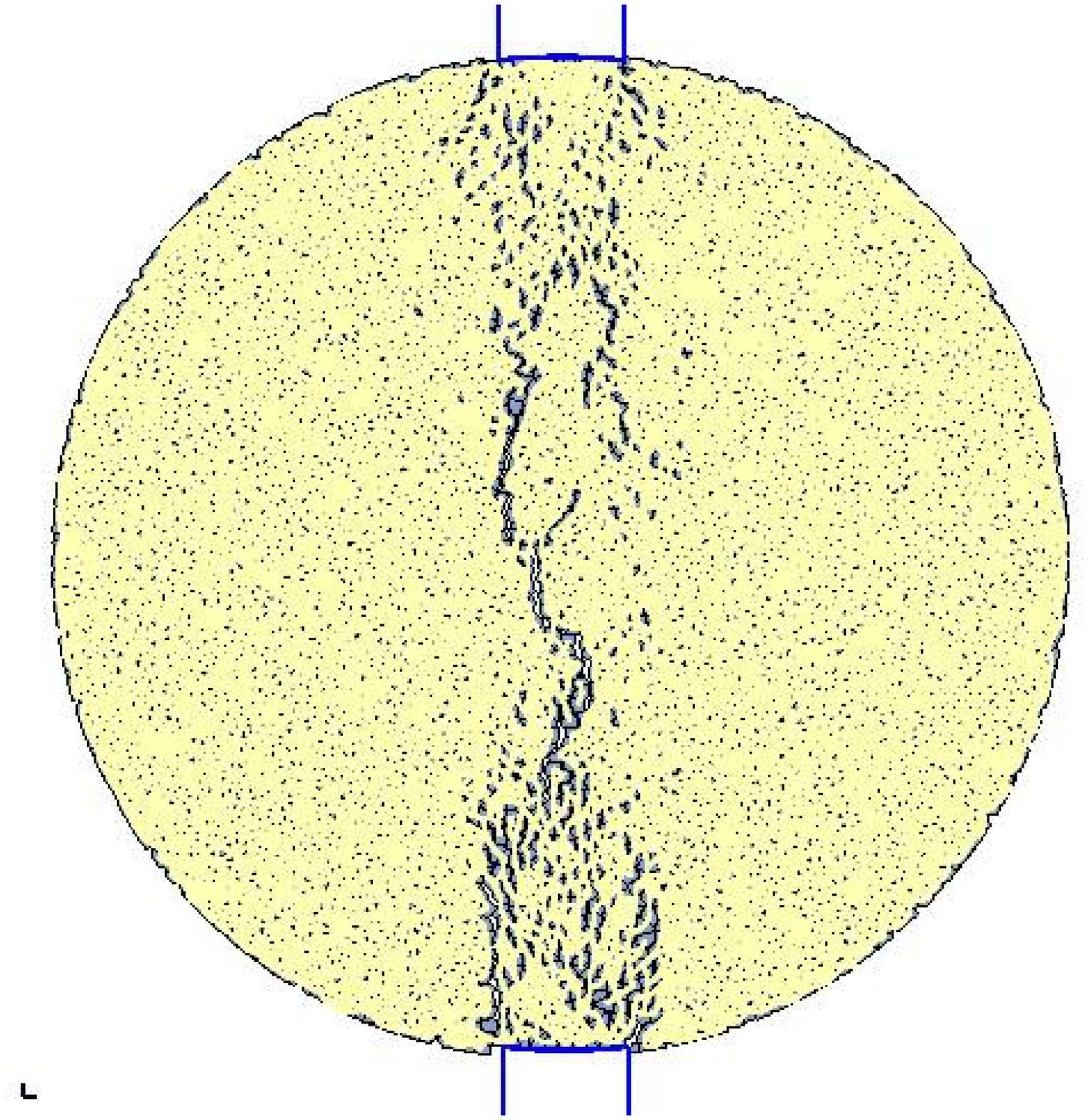}   &% t = 256
	\includegraphics[width=4.4cm,bb=0 0 555 555]{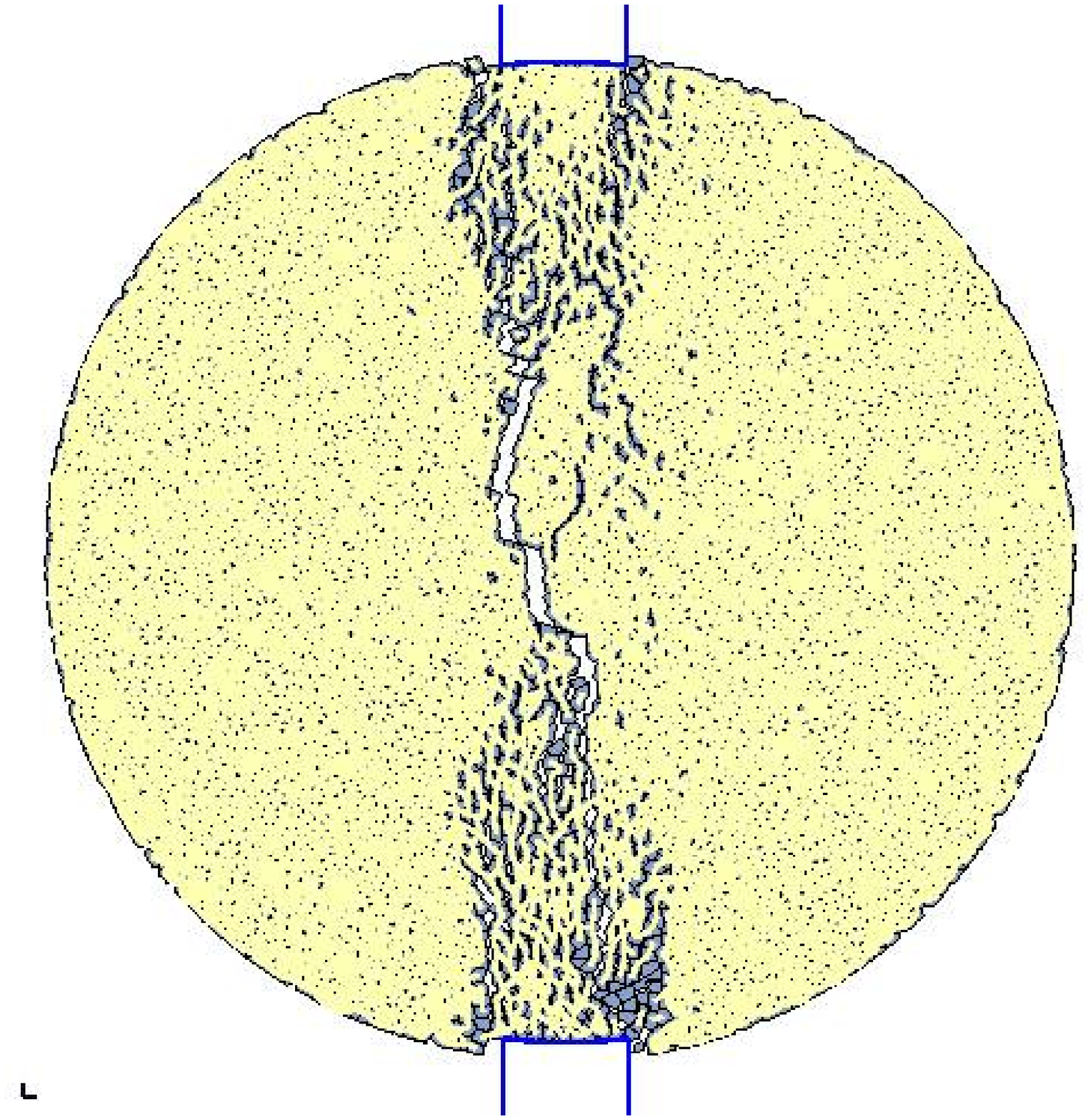} \\ [0.4cm] %t = 283
\end{array}$

\end{center}
	\caption{{\it (Color online)} Fracture of a disk-shaped solid
subjected to a constant diametric load. Snapshots of the evolving
damage. Simulations were carried out with the parameter values:
$\sigma/\sigma_c = 0.8$, $f_0=100$, and $\tau = 1300$. For further
parameters see Table \ref{tab:params}.}  
	\label{timeevol}
\end{figure}
Since usually the fracturing of materials is highly dependent on the
structural environment in which it is studied, one needs to reproduce
the experimental geometry as accurately as possible. Starting from the
Voronoi construction of a square, one cuts a disk-shaped specimen
by removing those polygons outside the disk region and reshaping those
at the border.  
The loading conditions for the numerical specimen are shown in Fig.
\ref{geometry}. This specimen is loaded in diametral compression by
non-deformable platens, which are modeled by introducing two new
polygons with the same physical properties of the loaded material. The
load platens are cut with the same curvature of the disk. The ratio
between the width of the load platen $b$ and the diameter of the disk
$D$ is $b/D=0.125$, as in the experiments of Ref. \cite{hans:424}. In
the simulations presented in this paper each disk 
specimen is composed of 5070 mesoscopic elements. 

\section{Results and Discussion}

\subsection{Fracture development}
In the numerical experiments the sample is loaded with a constant
external stress $\sigma$. In order to avoid undesirable large elastic
waves, the load applied to the opposite platens is slowly increased
from zero. When the load reaches the value $\sigma$, the numerical
sample is allowed to equilibrate during thousands of time steps
including a small damping between the grains and friction according to
Coulomb's friction law \cite{hans:267, hans:263}. 
\begin{figure}
\begin{center}
	\includegraphics[width=8.2cm,bb=50 50 530 400]{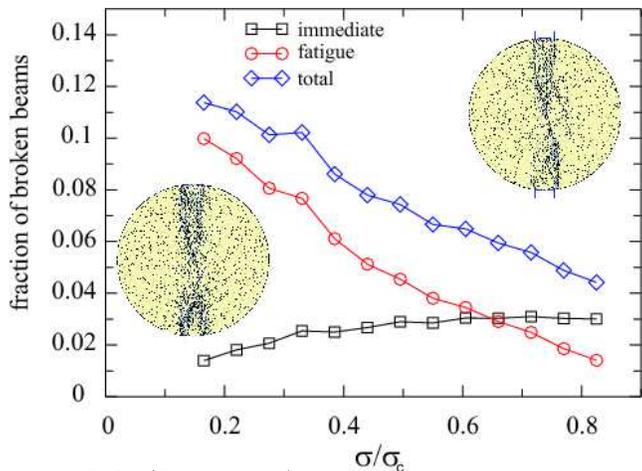}
\end{center}
	\caption{{\it (Color online)} Fraction of broken beams due to
immediate breaking (squares) and memory accumulation (circles) as a
function of $\sigma$, for $f_0=100$, $\tau=500$, together with the total fraction 
of broken beams (diamonds). The insets show the
final fracture pattern for $\sigma = 0.2$ (left) and $\sigma = 0.8$
(right).} 
	\label{fractions}
\end{figure}
Only after
equilibration the breaking rules are applied to the beams. With this
slow loading and long equilibration time, the vibrations of the sample
are drastically reduced compared to the case when a constant load is
applied instantaneously to the platens. The parameter values used in
the simulations are summarized in Table \ref{tab:params}. 

First we carried out computer simulations in order to determine the
quasi-static strength $\sigma_c$ of the model solid at the given
parameters. Then the damage and fracture of the disc was studied
varying the external load $\sigma$ below $\sigma_c$. 
In order to understand the failure process we studied in details the
evolution of the crack pattern during the simulations. Figure
\ref{timeevol} presents snapshots of the evolving system for a disk of
diameter 20 cm (approximately 80 polygons) and $\sigma/\sigma_c=0.8$. In the initial stage of
fracture, shown in Fig. \ref{timeevol}(a), a number of cracks
appears in the regions situated near the border of the platens
contacts. Because of the finite width of the platens, the stress field
induced in the sample is not purely uniaxial tensile stress along the
load direction as one might expect \cite{tsoungui:1999gm}, so the
cracks do not initiate from the middle of the specimen. In fact, the
cracks initiate from the border and coalesce to form wedges, as shown in Figure
\ref{timeevol}(b). The wedges extend up to approximately $0.25D$,
exhibiting a series of parallel vertical cracks, characteristic of
compressive tests \cite{hans:263}. As time evolves the wedges
penetrate the specimen (see Fig. \ref{timeevol}(c)) originating a
fracture in the center of the specimen, in a narrow region between the
load platens. This mechanism leads to the sudden and catastrophic
failure of the specimen (as shown in Figure \ref{timeevol} (d)). This failure
behavior agrees qualitatively well with previous experimental observations 
\cite{hans:424, vanMier:1999}. 
\begin{figure}
\begin{center}
$\begin{array}{c@{\hspace{0.0 cm}}c}
	\multicolumn{1}{l}{\mbox{\begin{large}(a)\end{large}}} &
 	\multicolumn{1}{l}{\mbox{\begin{large}(b)\end{large}}}  \\ [0.2cm]
	\includegraphics[width=4.4cm,bb=0 0 555 555]{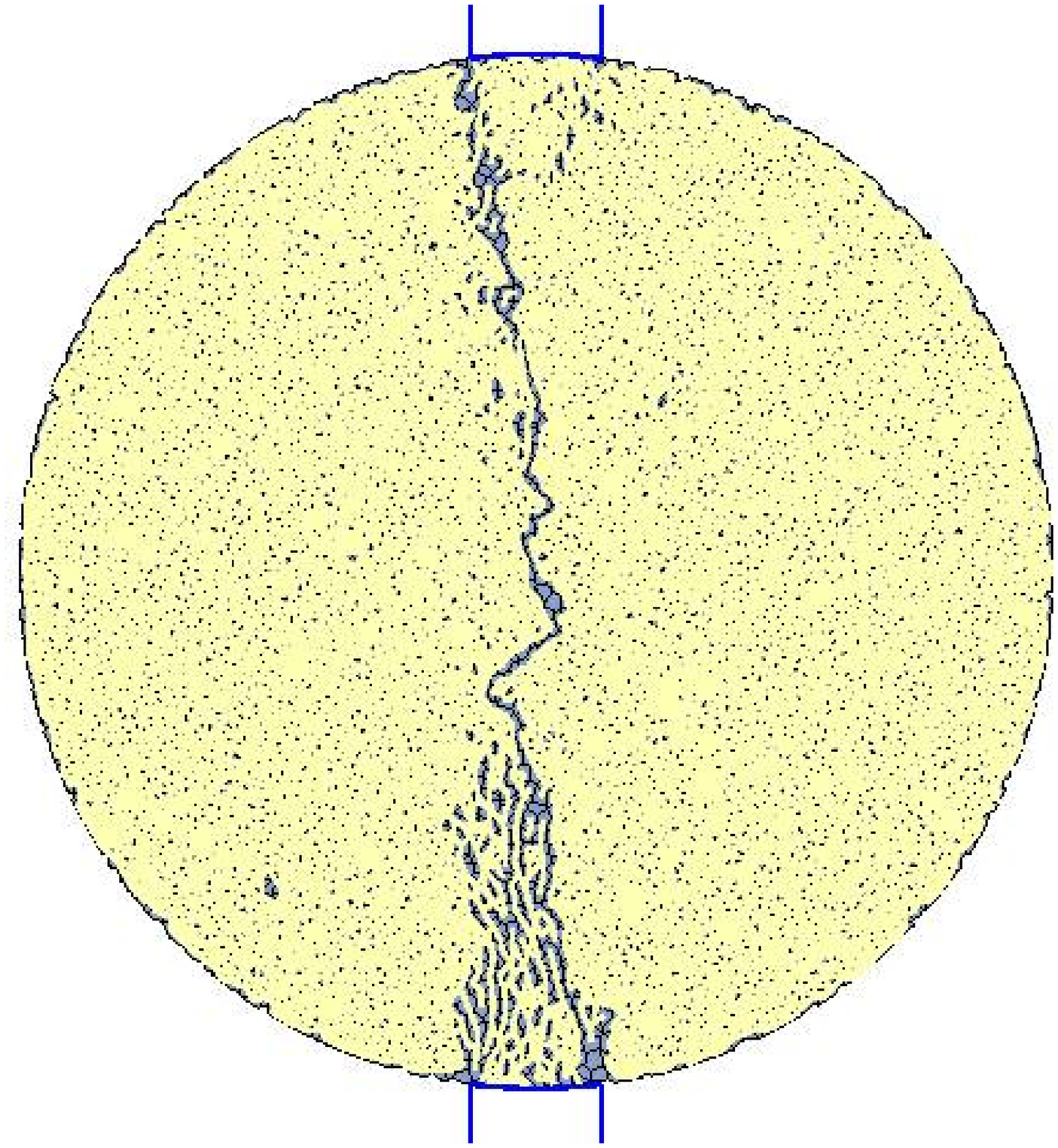}   &% f_0=1
	\includegraphics[width=4.4cm,bb=0 0 555 555]{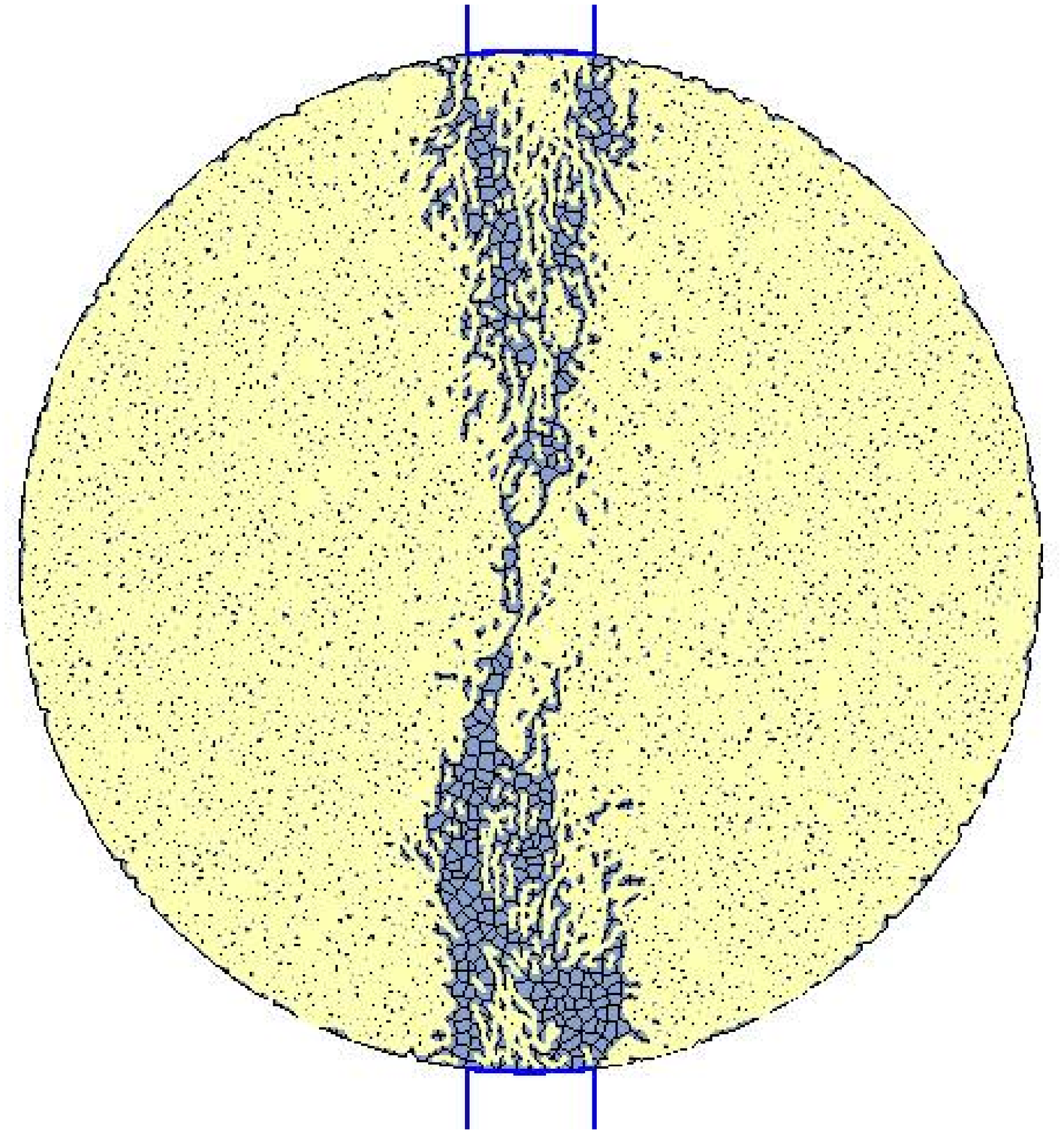} \\ [0.6cm] %f_0=100 
\end{array}$
\end{center}
	\caption{{\it (Color online)} Final fracture pattern with $f_0=1$ (a) and $f_0 =
100$ (b) obtained by computer simulations at the same load
$\sigma/\sigma_c=0.8$ and $\tau\to \infty$. At the higher value of the
memory factor $f_0$ more dispersed cracking occurs $(b)$, while for
low $f_0$ correlated crack growth can be obtained due to the dominance
of immediate breaking.}
	\label{f0influence}
\end{figure}

It is important to emphasize that without the time dependent damage
accumulation,
the system would not have macroscopic failure under a load $\sigma <
\sigma_c$. After some cracking events the disc would attain equilibrium
resulting in an infinite lifetime. However, according to the failure
criterion Eq.\ (\ref{pvalb}), intact elements accumulate damage which
results in beam breaking after a finite time even under a constant
local load. These beam breakings due to damage accumulation give
rise to load redistribution and stress enhancements on intact elements
triggering additional breakings. In this way, if healing is neglected, or
analogously $\tau \to \infty$ is set in Eq.\ (\ref{pvalb}), the
system would fail after a finite time at any small load value. 
One can conclude from Eq.\ (\ref{pvalb}) that the main role of the memory
factor $f_0$ is to control the relevant time scale of the
system, {\it i.e.} the lifetime of the specimen $t_f$ has a
dependence $t_f \sim 1/f_0$.

In order to analyze the importance of the accumulation of damage in
the fracture process, we have investigated the final fractions of the
beams that have been broken during the course of the simulations due
to the immediate breaking rule (first term in Eq.\ (\ref{pvalb})) and
those that have been broken due to damage accumulation (second term in
Eq.\ (\ref{pvalb})) for different applied loads. Figure \ref{fractions}
shows the fractions of the broken beams due to the different breaking
modes as a function of the applied load. For low load values damage dominates the failure of beams, while immediate 
breaking plays only a minor role. As the external load is increased, the fraction
of broken beams due to damage accumulation decreases monotonically,
while that due to immediate breaking increases and saturates. For
$\sigma \geq 0.65 \sigma_c$ the immediate breaking mode becomes dominant. 
It is interesting to note that the total fraction of broken elements is a decreasing
function of the load, implying that correlated crack growth dominates when $\sigma \rightarrow \sigma_c$

One of the most important features of our modeling approach is that it
accounts for the inhomogeneous stress field in the specimen and
provides also information on the microstructure and spatial
distribution of damage. The ageing of beams is less sensitive to the
details of the stress field in the specimen. That is why when damage
accumulation dominates the breaking of beams microcracks are more
dispersed in the stripe of the specimen between the loading platens. 
The insets of Fig. \ref{fractions} show the final fracture pattern
corresponding to $\sigma = 0.2 \sigma_c$ and $\sigma = 0.8
\sigma_c$. For $\sigma = 0.2$ (left), where the effect of fatigue is
more important, we observe more disordered cracks, while for $\sigma =
0.8$, where the immediate breaking process is more important, longer
cracks can be identified due to crack growth.

\begin{figure}
\begin{center}
	\includegraphics[width=8.2cm,bb=0 0 516 348]{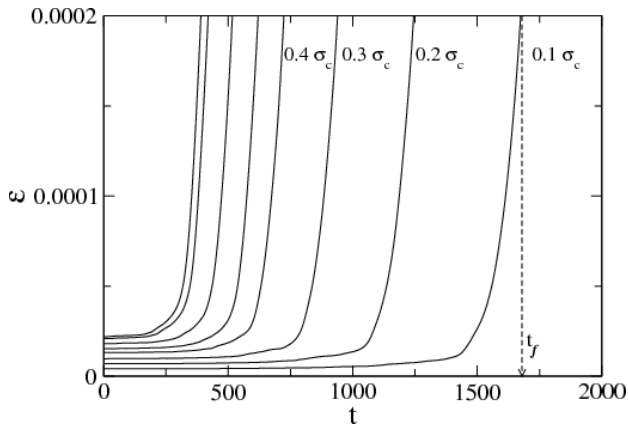}
	\caption{Deformation $\varepsilon$ of the disk as a function of
time $t$ for $f_0 = 100$ and $\tau = \infty$ varying the external
load $\sigma$. Macroscopic failure is preceeded by an acceleration of
the deformation. The vertical dashed line indicates the lifetime $t_f$
of the system. Note that $t_f$ decreases with increasing $\sigma$.} 
	\label{defxtime}
\end{center}
\end{figure}
In our model the importance of damage accumulation for the beam
breaking is controlled by the memory factor $f_0$ with respect to the
immediate failure of elements. Figure \ref{f0influence} demonstrates
crack patterns obtained at the same load without healing $\tau \to
\infty$ varying the value of $f_0$. It can be observed that for large
values of $f_0$, even at a relatively high load $\sigma/\sigma_c =
0.85$, a large amount of 
distributed cracking occurs in the specimen. Correlated crack growth
can only be observed for low $f_0$ (see 
Fig.\ \ref{f0influence}a) where immediate breaking has dominance. 
These crack patterns are very similar to those originated due to memory effects
in thermal fuse networks \cite{sornette:1992}.
The effect of increasing the memory factor $f_0$ resembles the effect of decreasing the applied
load, in the sense that increasing $f_0$ has the same effect of
enlarging the lifetime of the sample, as can be inferred from
Eq.\ (\ref{pvalb}). 

\subsection{Macroscopic time evolution}

On the macroscopic level the time evolution of the system is
characterized by the overall deformation $\varepsilon$ of the disk and
the lifetime $t_f$ of the specimen. 
The total deformation of the specimen, $\varepsilon$,  is obtained
directly from the simulation by monitoring the position of the loading
platens. Figure \ref{defxtime} shows the behavior of $\varepsilon$ as a
function of time for different values of $\sigma$ with respect to the
tensile strength $\sigma_c$ of the simulated specimen. Due to
accumulation of damage, $\varepsilon$ increases monotonically until
catastrophic failure of the material occurs. Some fluctuations can be
seen in $\varepsilon$ before it starts increasing rapidly, due to the
natural randomness of the beam breaking. The lifetime $t_f$ of the
specimen is defined as the time at which the deformation $\varepsilon$
diverges. It can be observed that, by increasing $\sigma$, the lifetime
$t_f$ rapidly decreases and actually goes to zero as the external load
approaches the tensile strength of the material $\sigma_c$.

\begin{figure}
\begin{center}
	\includegraphics[width=8.2cm, bb=0 0 434 288]{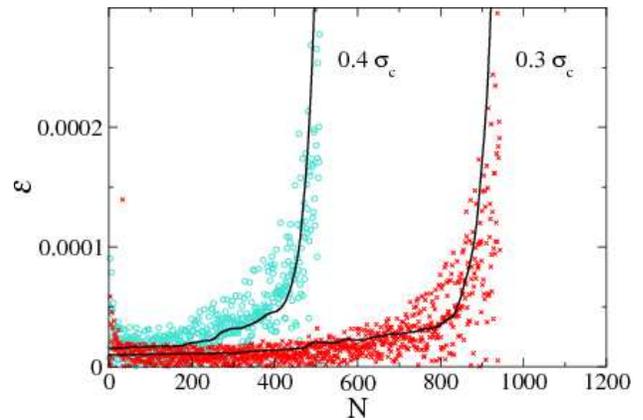}
	\caption{Fit of the experimental results of Ref.\
\cite{hans:424}. Using the parameter values $f_0 = 100$ and $\tau =
1300$ a good quality agreement is obtained with the experiments.}   
	\label{exprmtfit}
\end{center}
\end{figure}

For the purpose of fitting the experimental results, the most important
parameters of the model that can be tuned are the two threshold values
$\varepsilon_{th}$ and $\theta_{th}$ for the stretching and bending
deformation of the beams, as well as the memory factor $f_0$ and the
range of memory $\tau$. As we have pointed out, $f_0$ only sets the
time scale of the system, while the other three parameters have a
strong effect on the qualitative form of $\varepsilon(t)$ and
$t_f(\sigma)$. 

In the experiments of Ref.\ \cite{hans:424} the asphalt
specimen was subjected to periodic loading with a constant amplitude
monitoring the maximum deformation as a function of the number of
loading cycles and the number of cycles to failure. In our model a
constant load or a periodic load with a constant amplitude would have
the same effect, hence, we directly compare the simulation results to
the experimental findings.

Figure \ref{exprmtfit} shows a fit of the experimental results of
Ref.\ \cite{hans:424} to the simulations, obtained by relating
time to number of cycles and an appropriate choice of geometrical and
physical parameters for the mesoscopic elements and beams. We see that
the simulation and the experimental curves show excellent agreement.

\begin{figure}
\begin{center}

	\includegraphics[width=8.2cm]{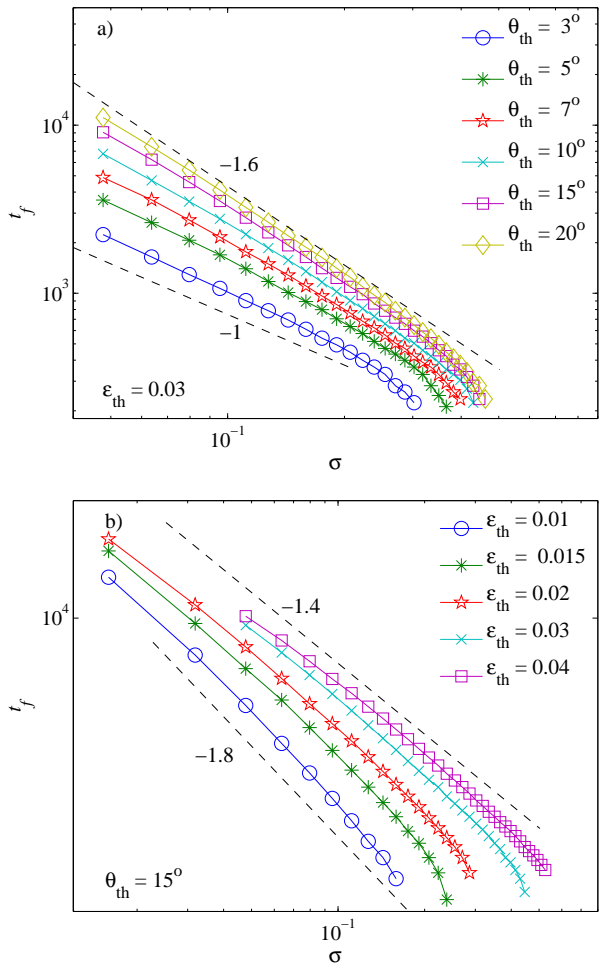}
	\caption{Lifetime as a function of the load $\sigma$ for different breaking thresholds.
	In (a) the stretching threshold is kept fixed  $\varepsilon_{th} = 0.03$ and we vary the bending angle threshold $\theta_{th}$, while in (b) $\theta_{th} = 15^{\rm o}$ and the lifetime is calculated for different values of  $\varepsilon_{th}$ .}
	\label{BreakingModes}
\end{center}
\end{figure}

\begin{figure}
\begin{center}

	\includegraphics[width=9.0cm]{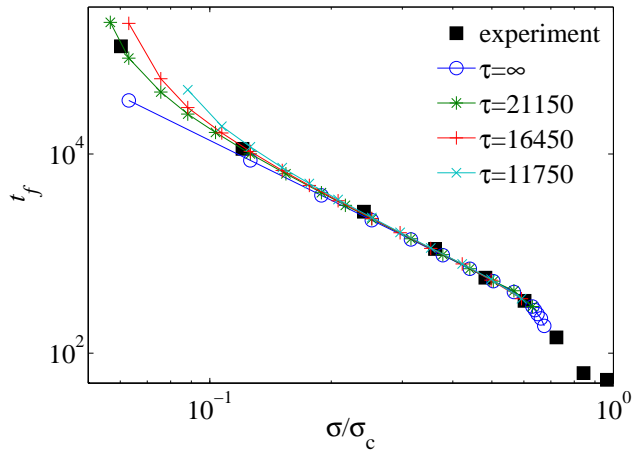}
	\caption{Lifetime as a function of the load
$\sigma/\sigma_c$. The square filled symbols correspond to
experimental results, the open circles correspond to $\tau=\infty$, and the $*$, $+$ and $\times$ symbols to $\tau =
	21150$, $16450$ and $11750$, respectively.} 
	\label{lifetime_b}
\end{center}
\end{figure}

In order to examine the consequences of the predominance of stretching or bending breaking modes
we performed a number of simulations with different threshold values. The dependence of the lifetime
on the external applied load is shown in Figure \ref{BreakingModes}, for $\tau = \infty$ and 
different values of $\varepsilon_{th}$ and $\theta_{th}$.
The simulations show that for external loads $\sigma \rightarrow
\sigma_c$, for all curves the lifetime decreases rapidly and
approaches the immediate failure of the specimen. For intermediate
values of applied load, $t_f$ exhibits a power law behavior, in
accordance with the Basquin law of fatigue \cite{basquin:1910} 
\begin{equation} 
t_f \sim \left( \frac{\sigma}{\sigma_c}\right)^{-\gamma}, \label{basquin}
\end{equation}
where the value of the exponent $\gamma$ can be controlled by the
breaking parameters $\varepsilon_{th}, \theta_{th}$. % and the range of memory $\tau$.
We can see in Fig. \ref{BreakingModes}a that for a fixed value of $\varepsilon_{th} = 0.03$
the Basquin exponent varies from $\gamma = -1.1 \pm 0.1$ for $\theta_{th} = 3^{\rm o}$ (bending breaking mode dominance) to
$\gamma = -1.6 \pm 0.1$  as we increase $\theta_{th}$ to $20^{\rm o}$, hence increasing the influence of the stretching 
breaking mode. Obviously, the value of $\sigma_c$ also increases with increasing $\theta_{th}$, since one of the breaking 
modes gets completely suppressed.
Similarly in Fig. \ref{BreakingModes}b, with a fixed value of $\theta_{th} = 15^{\rm o}$  the  Basquin exponent varies from $\gamma = -1.4 \pm 0.1$ for $\varepsilon_{th} = 0.04$ to $\gamma = -1.8 \pm 0.1$ for $\varepsilon_{th} = 0.01$, 
in which case there is a clear predominance of the stretching breaking mode. 

In order to obtain the measured value $\gamma = 2.0 \pm
0.1$ the breaking parameters had to be set such that the stretching
mode of Eq.\ (\ref{pvala}) dominates the breaking, {\it i.e.}
$\varepsilon_{th}=0.01$ and $\theta_{th}=20^{\rm o}$ were used which
implies that the bending mode has only a minor role in breaking.
The upturn of the curves of $t_f$ for small loads in Fig.\
\ref{lifetime_b} indicates the emergence of the fatigue limit
$\sigma_l$ in the system below which $\sigma < \sigma_l$
the specimen only suffers partial failure and has an infinite
lifetime. Figure \ref{lifetime_b} also illustrates that the value of
$\sigma_l$ is determined by the range of memory $\tau$, since healing
has only a dominating effect on the time evolution of the system when $\tau$ is comparable
to the lifetime of the specimen measured without healing ($\tau \rightarrow \infty$). Hence, it follows 
from the Basquin law, Eq.\ref{basquin}, that the fatigue limit $\sigma_l$ scales with $\tau$ as $\sigma_l \sim \tau^{-\frac{1}{\gamma}}$. 
The good quantitative agreement
between simulations and experiment obtained for $\tau = 21150$ indicates 
that the stretching breaking mechanism is more important for this particular material.
This is reasonable since the polymer binding in asphalt can not sustain
torsion.

\section{Conclusions}
We carried out a computational study of the fatigue failure of
disordered materials occurring under a constant external
load. A two-dimensional discrete element model was extended to capture
microscopic failure mechanisms relevant for the process of fatigue. As
a specific example, we considered experiments on asphalt where damage
recovery in the form of healing is known to play a crucial role for
the long term performance of the material. 

The breaking of beams is caused by two mechanisms: immediate breaking
occurs when the failure thresholds of stretching and bending are
exceeded. Intact beams undergo a damage accumulation process which is
limited by the finite range of memory (healing). The relevant
parameters of the model to fit the experimental results are the
breaking thresholds of stretching $\varepsilon_{th}$ and bending
$\theta_{th}$ and the range of memory $\tau$. We carried out a large
amount of computer simulations to study the time evolution of the
system at the macro- and micro-level. 
We demonstrated that at the
micro-level failure of beams due to damage is responsible for the
diffuse crack pattern, while correlated crack growth is obtained when
immediate breaking has dominance. The model provides a good
quantitative agreement with previous experimental findings on the
fatigue failure of asphalt, {\it i.e.} varying only three parameters
$\varepsilon_{th}$, $\theta_{th}$, and $\tau$ of the model we could
reproduce the deformation-time diagram and the Basquin-law of
asphalt obtained in experiments. Very interestingly we found that the
value of the Basquin exponent of the model is controlled by the
relative importance of the stretching and bending modes in beam
breaking. 
The fatigue limit $\sigma_l$ is obtained as a threshold
value of the external load below which the specimen does not suffer
macroscopic failure and has an infinite lifetime. In the framework of
our model $\sigma_l$ is solely determined by the range memory $\tau$
over which local loads contribute to the total accumulated damage in
the system. 

% We determined the scaling structure of the deformation-time
% characteristics of the system obtained at different external
% loads. We showed that one of the scaling exponents coincides with the
% Basquin exponent of the material, while the other one has value 1
% independent on material parameters. The scaling structure of
% $\varepsilon(t)$ is an interesting prediction for future experimental
% tests. 
\section{Acknowledgments}

We thank the Brazilian agencies CNPq, FUNCAP and CAPES for financial
support. H.\ J.\ Herrmann is grateful to Alexander von Humboldt Stiftung
and for the Max Planck Prize. F.\ Kun was supported by OTKA M041537,
and T049209.

\bibliography{matsci}
\end{document}